\documentstyle[prl,aps,epsf]{revtex}
\begin{document}
\advance\textheight by 0.5in
\advance\topmargin by -0.25in
\draft

\twocolumn[\hsize\textwidth\columnwidth\hsize\csname@twocolumnfalse%
\endcsname

\title{ 
Thermal transport in a Luttinger liquid}
 
\author{C.L. Kane}
 
\address{Department of Physics, University of Pennsylvania\\
Philadelphia, Pennsylvania 19104
}
\author{Matthew P.A. Fisher}
\address{Institute for Theoretical Physics, University of California\\
Santa Barbara, CA 93106-4030
}

\date{\today}

\maketitle

\begin{abstract}
We study thermal transport in a one-dimensional (1d)
interacting electron gas, employing the Luttinger liquid model. 
Both thermal conductance and thermopower are analyzed for a pure
1d gas and with impurities.  
The universal ratio of electrical to thermal conductance
in a Fermi-liquid - the Wiedeman-Franz law - is
modified, whereas the thermopower is still linear in temperature. 
For a single impurity the Lorenz number is
given by $L(T \rightarrow 0) = 3L_0/(2g+g^2)$ - with $L_0$ the Fermi liquid value -
and the conductance
$1/2 < g < 1$.
For $g<1/2$ the Lorenz number {\it diverges} as
$T \rightarrow 0$.  Possible relevance to thermal transport in conducting polymer
systems is discussed. 
\end{abstract}
\vskip -0.5 truein
]

The Wiedemann-Franz law, which relates the thermal 
and electrical conductivity ($\kappa$,$\sigma$) of metals, 
played a central role in the historical development of 
the quantum theory of solids.
The Lorenz number, $L = \kappa/\sigma T$,
originally computed within classical Drude theory,
gave fortuitous agreement with experiment due to cancelling errors. 
The quantum theory corrected the errors, and improved the
agreement.  For non interacting electrons
Chester et. al.\cite{Chester} showed
that the Lorenz number is given exactly by $L_0 = (\pi^2/3)(k_B/e)^2$,
for {\it arbitrary} impurity scattering strength.
In the 1980's Castellani et. al.\cite{Castellani} 
argued that this universal value was robust
even with inclusion of electron interactions,
provided the system remained metallic.
Thus, a universal Lorenz number appears to be
a defining characteristic of the Fermi liquid phase.

In recent years there has been tremendous interest in conducting phases
which are {\it not} Fermi liquids.  A paradigm for these are 
1d interacting electron gas models,
which exhibit a non Fermi liquid phase even for weak interactions
\cite{Luttinger}.  The resulting 
Luttinger liquid phase is characterized by a dimensionless
conductance, $g$, which controls various power laws, such
as the singularity in the momentum distribution function.
The resurgence of interest in the 1d Luttinger liquids
stems both from the recent ability to lithograghically pattern
true one-channel quantum wires\cite{wire}\cite{KF} and from the realization that 1d edge states
in the fractional quantum Hall effect are Luttinger liquids\cite{Edge}.
Other non-Fermi liquid phases arise in quantum impurity
problems\cite{Ludwig}, such as the multi-channel Kondo
model which is possibly relevant to heavy Fermion materials.
Bulk 2d non-Fermi liquid phases have also been suggested 
in compressible Hall fluid phases\cite{HLR}
and in the cuprate superconductors\cite{marginal}.

It is natural to anticipate that thermal transport
in such non-Fermi liquid phases
will be qualitatively different, and might help
characterize and distinguish them experimentally.  In this paper, 
we consider in detail thermal
transport in the 1d Luttinger liquids.
We show that the Lorenz number can be substantially modified
from its Fermi liquid value, $L_0$.  
The thermopower, $Q$, 
on the other hand, shows characteristically metallic behavior,
$Q =  c  T$.  As in conventional metals,
the coefficient $c$ is non-universal, depending on the curvature of the
energy bands and the energy dependence of the scattering
rates.

While thermal transport measurements in quantum wires and quantum
Hall samples are undoubtedly extremely challenging, a remarkable
recent experiment has demonstrated the feasibility
of such experiments\cite{Vanhouten}.  Thermal transport 
measurements in bulk quasi 1d
samples, such as conducting polymers, are much easier, 
but 3d
crossovers may tend to complicate the analysis.  

{\it Pure Luttinger liquid}:  
We begin with a model for an interacting spinless 1d electron gas
in the absence of any impurities,
which has a bosonized Hamiltonian density
\begin{equation}
{\cal H}_0 = \pi v_0 (N_+^2 + N_-^2 + 2\lambda N_+N_-) .
\end{equation}
The right and left moving electron densities, $N_{\pm}$,
satisfy Kac-Moody commutation relations:
\begin{equation}
[N_{\pm}(x),N_{\pm}(x^\prime)] = \pm (i/2\pi) 
\partial_x \delta(x-x^\prime)  .
\end{equation}
The interaction term mixes right and left movers, but
can 
be shifted away as usual by defining new fields
\begin{equation}
N_{\pm}= [g(n_+ + n_-) \pm (n_+ - n_-)]/2g
\end{equation}
with $\lambda = (1-g^2)/(1+g^2)$.  
In terms of $n_\pm$ the Hamiltonian decouples
into right and left moving sectors:
\begin{equation}
{\cal H}_0 = {\cal H}_0^+ + {\cal H}_0^- = (\pi v/g)(n_+^2 + n_-^2) 
\end{equation}
with renormalized velocity, $v=(2gv_0 )/(1+g^2)$.
The new fields also satisfy a Kac-Moody algebra,
\begin{equation}
[n_{\pm}(x),n_{\pm}(x^\prime)] = \pm (ig/2 \pi ) 
\partial_x \delta(x-x^\prime) .
\end{equation}

Consider now transport
in an ideal Luttinger liquid.
Intitally, we ignore additional
anharmonic interaction terms
(eg. $n_+^2n_-$) which couple the right and left moving modes in (4).
(For chiral quantum Hall edge states, these will be absent.)
It then suffices 
to consider only a single 
right moving channel, $n=n_+$.
Such an ideal chiral channel can be characterized by transport
coefficients ${\cal L}^{ij}$, which relate changes in the
electrical and thermal currents to changes
in the chemical potential, $\mu$, and temperature $T$.
These coefficients are equivalent to  
Landauer two-terminal transport coefficients\cite{Imry}, defined 
with ``ideal" reservoirs.
For ideal quantum wires, these coefficients are 
not measured directly, since the contacts do not couple
selectively to right and left moving modes.  However, they
can be measured directly for quantum Hall edge states.

The charge density in a chiral channel is conserved
by (4), and satisfies
$\partial_t n + \partial_x J =0$ with an
electrical current $J=vn$.  Changing the chemical potential, $\mu$,
alters the electrical
current.  Balancing the $n^2$ energy in (4) with a $-\mu n$
term, gives $\Delta J=(g/2\pi)\Delta\mu$, or upon restoring units an electrical
conductance, $G = {\cal L}^{11} =ge^2/h$.  

Heat carried by a chiral channel is likewise conserved
by (4).  The continuity equation
$\partial_t n_Q + \partial_x J_Q =0$ is satisfied
by the thermal energy density $n_Q=(\pi v/g) n^2$
and thermal current $J_Q=vn_Q$.
The thermal energy 
at temperature $T$ can be expressed in terms of the
chiral Luttinger modes
as
\begin{equation}
n_Q = \int_0^\infty {{dk} \over {2\pi}} \omega_k b_{\omega_k}
\end{equation}
with $b_\omega = (e^{\beta \omega} -1)^{-1}$ and $\omega_k = vk$.
This gives $J_Q = (\pi^2/6)(k_B T)^2/h$, and leads to a
``quantized" thermal conductance, 
$K = {\cal L}^{22} = \partial J_Q/\partial T = (\pi^2/3) k_B^2 T/h$.

For an ideal Luttinger liquid we
can then define a``two-terminal" Lorenz number
\begin{equation}
L_{\rm ideal} = K/TG = L_0/g.
\end{equation}
For $g=1$, we recover the Fermi-liquid
value $L_0 = (\pi^2 /3)(k_B/e)^2$.
With repulsive interactions ($g<1$) the Lorenz number
is larger.

The off-diagonal transport coefficient 
${\cal L}^{12}= \partial J/ \partial T$, which 
determines the thermopower, is zero within
the present model, due to the implicit linearization of the electronic
bandstructure near the Fermi energy.  The effects of dispersion
can be included via the third order interaction term,
${\cal H}_{\rm int} = A n^3$, which
is normally ignored because it is formally
``irrelevant".
The coefficient $A$ is proportional to the change in Fermi velocity
with chemical potential, $dv/d\mu$.  
The resulting thermopower, $Q = {\cal L}^{12}/{\cal L}^{11}$, 
is linear in temperature,
\begin{equation}
Q = - (\pi^2 k_B^2 /3  g e v) (dv/d \mu ) T .
\end{equation}

In quantum wires, anharmonic interactions ignored above
will couple the right and left moving modes.
The right and left
moving thermal currents will no longer be independently
conserved.  However, in a translationally
invariant system, thermal currents cannot fully
relax due to constraints of momentum conservation.
Such anharmonic interactions might nevertheless effect the
value of $K$. 
Umklapp processes would allow a decay of thermal current,
but freeze out at low temperatures.
In any event, impurity backscattering will
dominate these interaction effects in
the thermal resistance of real quantum wires.

{\it Single Impurity}:
We now consider a single impurity in an otherwise ideal
Luttinger liquid, as a first step toward inclusion of many impurities.
(A single impurity is also relevant to point contact  
experiments in the quantum Hall effect.)
A weak potential scatterer at the origin can be modelled by adding a term
to the Hamiltonian,
\begin{equation}
{\cal H}_{Back} = - t_B \cos(\phi_+ - \phi_-) \delta(x)  ,
\end{equation}
where $t_B$ is the amplitude for $2k_F$ 
electron backscattering.  This process has been expressed
in terms of  
the boson fields $\phi_{\pm}$, related to the
densities $n_{\pm} = \pm \partial_x \phi_{\pm}/2\pi$.
The operator $\exp(i\phi_+)$
creates an excitation with fractional charge, $ge$.
Thus each backscattering process reflects fractional
electron charge.

An impurity which strongly backscatters can alternatively be modelled 
as a tunnel junction
between two de-coupled semi-infinite Luttinger liquids\cite{KF}.
In this case, the chiral density $n_+$ can be taken to
describe the right and left
moving pieces of one semi-infinite Luttinger liquid.
The appropriate term which tunnels an electron (charge e) through
the junction is then,
\begin{equation}
{\cal H}_{tunn} = - t \cos \left( (\phi_+ - \phi_-)/g \right) \delta(x) .
\end{equation}

To proceed we follow Fendley et. al.\cite{Fendley} and first
define new fields which propogate in the same direction: $\phi_1(x) = \phi_+(x)$ and $\phi_2(x) = \phi_-(-x)$,
and associated densities, $n_j = \partial_x \phi_j$ with $j=1,2$.
One can then define commuting even and odd densities,
$n=n_+ - n_- = \partial_x \phi /2\pi$ and $N=n_+ + n_-$.
The full Hamiltonian with backscattering present factorizes,
\begin{equation}
{\cal H} = (\pi v/2g)(n^2+N^2) - t_B \cos\phi \delta(x) .
\end{equation}

The backscattered electrical current is given by
\begin{equation}
J = \int_x \partial_t n/2 =  g t_B \sin\phi(x=0)  ,
\end{equation}
where the second equality follows from commuting $n$ with the Hamiltonian.
For the case of a tunnel junction, the tunnel current is
$J=t\sin(\phi(x=0)/g)$.
Similarly, the backscattered thermal current can be written,
\begin{equation}
J_Q = \int_x \partial_t ({\cal H}_0^+ - {\cal H}_0^- )/2 = (\pi v/g) N(x=0) J  ,
\end{equation}
where again the time derivatives are evaluated by commuting with the
Hamiltonian.  This form also holds for the tunnel junction.
Notice that the thermal current has been decomposed into a product
of two commuting contributions:  The even density $N$, and the
electrical current $J$ which depends only on the odd boson.
This remarkable simplification, enables us to derive 
an expression relating the thermal and electrical conductances.

To this end consider the current correlation function,
\begin{equation}
P^R(t) = i \Theta(t) <[J(t),J(0)]> = i\Theta(t) (P^>(t)-P^<(t))
\end{equation}
from which the electrical conductance follows:
\begin{equation}
Re G(\omega) = {1 \over \omega} Im P^R(\omega) = 
{1 \over {2\omega}} (P^>(\omega)-P^<(\omega))  .
\end{equation}
Denoting the corresponding correlators for the thermal current, $J_Q$,
with a subscript $Q$, the thermal conductance can be obtained
from:
\begin{equation} 
K = \lim_{\omega \rightarrow 0} {1 \over {\omega T}} Im P_Q^R(\omega)
= {1 \over {2T^2}} P^<_Q(\omega = 0) ,
\end{equation}
where the latter equality follows
upon using the detailed balance relation,
$P^>_Q(\omega) = \exp(\beta \omega) P^<_Q(\omega)$. 

The relation between the thermal and electrical current operators (13),
allows us to relate their respective correlation functions:
\begin{equation}
P_Q^<(t) = (\pi v/g)^2  D_2^<(t) P^<(t)
\end{equation}
where $D_2$ is the even boson density-density ($N-N$) correlation function.
Using the fact, $D_2^<(t)=D_2^>(-t)$, allows one to express the thermal 
conductance as:
\begin{equation}
K = {1 \over 2} ( {{v \pi} \over {gT}} )^2 
\int {{d\omega} \over {2 \pi}}  D^>_2(\omega) P^<(\omega)  .
\end{equation}
The function $D_2^>$ can be readily extracted since 
$\cal H$ in (11) is quadratic in $N$, giving $D_2^>(\omega) =
(g/\pi v^2) \omega e^{\beta \omega} b_\omega$. 
In addition using (15), along with detailed
balance, relates $P^<$ to the electrical
conductance: $P^<(\omega) = 2\omega b_\omega Re G(\omega)$.
We thereby obtain our final expression relating the
thermal and electrical conductances through the impurity:
\begin{equation}
K = {1 \over {8gT^2}} \int d\omega {{\omega^2 Re G(\omega)} \over
{\sinh^2(\omega/2T)}}   .
\end{equation}

In the absence of any backscattering, $Re G(\omega) = g/2\pi$,
which gives the pure result $K = (\pi^2/3)k_B^2T/h$.
In the limit of strong backscattering, $K$ can be obtained
from (18) by
calculating $P^<(\omega)$ perturbatively in powers of
$t$, the electron amplitude to tunnel through the junction.
To leading order, $P$ is the correlation function of the current,
$J=t\sin(\phi/g)$, evaluated with the free odd boson Hamiltonian
(ie. (11) with $t_B=0$).  One finds $K = c t^2
T^{2/g-1}$,  with $c$ a non-universal constant depending on a
short-time cut off.
This same constant also enters the electrical conductance,
and so drops out in the Lorenz ratio, which in this limit is found to be:
\begin{equation}
L =  3L_0 /(2g+g^2) .
\end{equation}
For non-interacting electrons ($g=1$) this reduces
to the Fermi-liquid value, but with repulsive interactions ($g<1$),
is larger.  

As one lowers the temperature, the Lorenz number crosses over
from the pure value, $L=L_0/g$ to the strong backscattering value (20).
For the special case of $g=1/2$, an expression for
this crossover can be obtained explicitly.  In particular, when $g=1/2$,
a closed form expression for the a.c. conductance follows
from the exact solution\cite{KF}:
\begin{equation}
Re G(\omega) = {g \over {2\pi \omega}} \int dE (f_E - f_{E+\omega})
{{E^2} \over {E^2 + T_B^2}}  
\end{equation}
with $f_\omega = (e^{\beta \omega} +1)^{-1}$.
Here $T_B$ is a crossover temperature scale, $T_B \sim t_B^{1/(1-g)}
\sim t_B^2$ (for $g=1/2$).  Together (19) and (21) allow one to compute
the Lorenz number for arbitrary $T/T_B$.  For $T>>T_B$ one finds
$L=2L_0$ in agreement with the pure result (7), but 
for $T<<T_B$, the result is $L=18L_0/5$ - a factor of 3/2 {\it larger}
than the strong backscattering result (20).  This discrepancy can be
traced to an irrelevant operator ignored in the 
perturbative
calculation leading to (20), but included implicitly in the $g=1/2$
crossover.
Specifically,  consider
a perturbation coupling the electron densities across the junction
in the strong backscattering limit:
\begin{equation}
{\cal H}_{pert} = a \delta(x) n_+ n_-  .
\end{equation}
This term does not transfer charge across the junction, but does transfer
energy and so contributes to thermal conduction.
Moreover, it feeds into the a.c. electrical 
conductance \cite{Gomez} as $Re G(\omega) \sim a^2 \omega^2$. 
Insertion in (19) then gives a contribution
to the thermal conductance varying as
$K \sim a^2 T^3$, which must be added
to the $T^{2/g-1}$ term coming from electron tunnelling.

\begin{figure}
\epsfxsize=3.5in
\epsffile{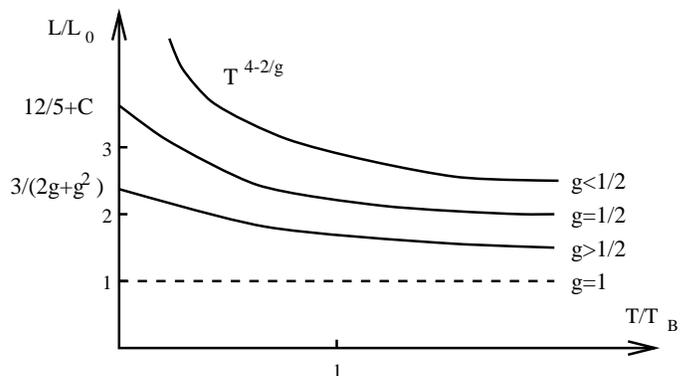}
\caption{
Lorenz number $L$ for
transport though a single impurity versus temperature, for several values of g.
}
\end{figure}

Three cases should then be distinguished, as sketched in the Figure.
For $g > 1/2$ the electron tunnelling term dominates
the thermal transport at low temperatures, and (20) is correct
as $T \rightarrow 0$.
For the soluble case $g=1/2$ both processes vary as $T^3$
and contribute to the  
Lorenz number as $T \rightarrow 0$.
Since the coefficient ``a" depends on details of the junction, a non-universal
Lorenz number is predicted in this case. However, 
since the contribution
to $K$ from (22) is positive (proportional to $a^2$), the Lorenz number 
should be bounded below    
by (20): $K = 12/5 + C$ with $C>0$,
consistent with the exact solution.
Finally, for $g< 1/2$, the interaction term (22) dominates
$K$ at low temperatures and
the Lorenz number {\it diverges} as $T \rightarrow 0$:  
\begin{equation}
L \sim (a^2 T^3)/T^{2/g-1} \sim a^2 T^{4-2/g}  .
\end{equation}
Physically, the inclusion of strong electron
interactions enables heat to be transported across the junction
much more readily than charge.

To compute the thermopower associated with transport through the impurity
it is necessary to include in the Hamiltonian (11)
terms which break particle/hole symmetry.
In addition to bulk cubic interactions
arising from dispersion, another local term
is of the form $Ncos(\phi)\delta(x)$, which
arises from an energy dependence of
the matrix element $t_B$.
One finds a thermopower linear in temperature, $Q = c T$,
with non-universal coefficient.

The present results can readily be generalized to include
the electron spin degree of freedom.
Assuming $SU(2)$ spin
symmetry, the Luttinger liquid can be characterized by
a dimensionless charge conductance $g_\rho$, which is equal
to $2$ for non-interacting electrons\cite{KF}.  For $g_\rho>2/3$ the Lorenz 
number $L$ is found to cross over from $L_0 (2/g_\rho)$
to $3L_0 (g_\rho + 2)^2/(8g_\rho(g_\rho+1))$ as one lowers the temperature
and scales from
the weak to strong backscattering regimes.
For $g_\rho < 2/3$, $L$ 
diverges as $T^{3-2/g_\rho}$ in the low temperature limit.

{\it Many Impurities}:
When many impurities are present in a repulsively interacting
1d Luttinger liquid, the conducting state is unstable to localization.
However, in systems with sufficiently dilute
but strong scatterers, there will be a range of temperatures
over which the 
above single impurity results should be observable.
Specifically, consider a model of many 1d conductors,
which are coupled electrically by dilute weak-links,
with a typical large separation $L$.  These weak-links would
serve as ``impurities" in the 1d transport.
Such a model might be appropriate for some conducting polymer systems,
in which the polymer backbones provides the 1d
conduction channel, and the weak-links
arise from tunnelling between different polymers at their ends.  
Applicability of this model requires the temperature
to be above the
tunnelling rate, $t_\perp$, between
parallel neighboring polymers.  Otherwise
the transport would no longer be 1d, but would take place coherently
across many chains.
Moreover, we require that tunnelling events across
successive weak-links
be incoherent,  so that $k_BT > \hbar v/L$, 
where $v$ is the 1d Fermi velocity.
In this temperature regime, the bulk electrical conductivity
should vary as a power law, $\sigma (T) \sim T^{2/g -2}$, before crossing
over at lower temperatures either into a localized regime or a bulk
metallic state.
Our single impurity results
imply that for $g>1/2$, the thermal conductivity
should also vary as a power law, $\kappa \sim T^{2/g-1}$,
with a universal non-Fermi liquid Lorenz number given by (20).
For $g< 1/2$ an even larger (and diverging) Lorenz ratio
is predicted.  

In some conducting polymer samples, the electrical
conductivity does vary as a power law with temperature\cite{Heeger}.
This power law has been interpreted\cite{Heeger}
as being in the vicinity of a bulk metal-insulator transition.
However, at a 3d Anderson localization transition
a temperature independent thermopower is predicted\cite{Barnes}, 
in contrast to the measured 
behaviour, $Q \sim T$.
In a model of 1d conductors with dilute weak-links,
a linear metallic thermopower would be expected.
The Lorenz number provides a further difference between these two
models.  At the
Anderson transition the Lorenz number is predicted 
to be supressed\cite{Barnes} below
the Fermi-liquid value $L_0$ by roughly 2/3, whereas our results
show an enhanced Lorenz number for 1d thermal transport through
dilute impurities.  It would be most interesting to measure thermal conductivity
in conducting polymer samples which exhibit power law 
electrical conductivities.

We thank D.S. Fisher,
A.J. Heeger, A.W.W. Ludwig, E.J. Mele, R. Menon and
S.M. Girvin for
clarifying conversations.  We are grateful to the National
Science Foundation for support.  M.P.A.F has been supported by
grants PHY94--07194, DMR--9400142
and DMR-9528578.  C.L.K. has been supported by grant DMR 95-05425.

\end{document}